\documentclass[aps,pra,amsmath,twocolumn,amssymb,footinbib,superscriptaddress]{revtex4}

\usepackage{amssymb,amsfonts,amsmath}

\usepackage{float}
\usepackage{color}

\usepackage[apple mac]{inputenc}
\usepackage[english]{babel}
\usepackage{times}
\usepackage{latexsym}
\usepackage{fancyhdr}
\usepackage{verbatim}
\usepackage{graphicx}
\usepackage{epsf}
\usepackage{subfigure}
\usepackage{bm}
\usepackage{epstopdf}

\definecolor{Nathanblue}{rgb}{0.,0.24,0.51}
\newcommand{\blue}{\color{Nathanblue}}

\def\be{\begin{equation}}
\def\ee{\end{equation}}
\def\ba{\begin{align}}
\def\enda{\end{align}}
\def\bi{\begin{itemize}}
\def\ei{\end{itemize}}
\def\bs#1{\boldsymbol{#1}}
\def\txt#1{\textrm{#1}}
\def\sf#1{\mathsf{#1}}

\def\Rb87{^{87}\rm{Rb}}						
\def\Li6{^{6}\rm{Li}}						

\begin{document}



\title{{\blue Realizing non-Abelian gauge potentials in optical square lattices: \\ Application to atomic Chern insulators}}

\author{N. Goldman}
\affiliation{Center for Nonlinear Phenomena and Complex Systems - Universit\'e Libre de Bruxelles , 231, Campus Plaine, B-1050 Brussels, Belgium}

\author{F. Gerbier}
\affiliation{Laboratoire Kastler Brossel, CNRS, ENS, UPMC, 24 rue Lhomond, 75005 Paris}
\author{M. Lewenstein}
\affiliation{ICFO -- Institut de Ci\`encies Fot\`oniques, Parc Mediterrani de la Tecnologia, 08860 Barcelona, Spain}
\affiliation{ICREA -- Instituci\'o Catalana de Recerca i Estudis Avan\c cats, 08010 Barcelona, Spain}

\begin{abstract}
We describe a scheme to engineer non-Abelian gauge potentials on a square optical lattice using laser-induced transitions. We emphasize the case of two-electron atoms, where the electronic ground state $g$ is laser coupled to a metastable state $e$ within a state-dependent optical lattice. In this scheme, the alternating pattern of lattice sites hosting $g$ and $e$ states depict a checkerboard structure, allowing for laser-assisted tunneling along both spatial directions. In this configuration, the nuclear spin of the atoms can be viewed as a ``flavor" quantum number undergoing non-Abelian tunneling along nearest-neighbor links. We show that this technique can be useful to simulate the equivalent of the Haldane quantum Hall model using cold atoms trapped in square optical lattices, offering an interesting route to realize Chern insulators. The emblematic Haldane model is particularly suited to investigate the physics of topological insulators, but requires, in its original form, complex hopping terms beyond nearest-neighboring sites. In general, this drawback inhibits a direct realization with cold atoms, using standard laser-induced tunneling techniques. We demonstrate that a simple mapping allows to express this model in terms of \emph{matrix} hopping operators, that are defined on a standard square lattice. This mapping is investigated for two models that lead to anomalous quantum Hall phases. We discuss the practical implementation of such models, exploiting laser-induced tunneling methods applied to the checkerboard optical lattice. 
\end{abstract}

\date{April 19th 2013}

\maketitle

\section{Introduction}

The discovery of the quantum Hall (QH) effect, in two-dimensional electronic systems subjected to large magnetic fields, revealed the existence of novel quantum phases: the topological insulating states \cite{Hasan:2010,Qi:2011}. Such phases are insulating in the bulk but exhibit current-carrying modes on the edge of the sample. These chiral propagating modes, whose energies are located within the bulk gaps, are \emph{protected by topology}. They can be formally related to the existence of non-trivial topological invariants -- Chern numbers -- associated with the band structure \cite{Kohmoto:1985,Hatsugai1993}. The existence of edge modes is guaranteed as long as the topology of the band structure does not change, irrespective of the details of the microscopic Hamiltonian. In the early age of quantum Hall physics, it was generally thought that this phenomenon could only emerge in two-dimensional systems featuring Landau levels, namely, in samples necessarily subjected to magnetic fields. Haldane showed in 1988 that this was not a necessary condition \cite{Haldane:1988}. To this aim, he introduced a simple lattice model exhibiting the quantum Hall effect without Landau levels, thereafter referred to as the \emph{anomalous} QH effect. In his seminal work \cite{Haldane:1988}, Haldane showed that quantum Hall phases were rooted in the breaking of time-reversal symmetry, which potentially increased the possibilities to access this phenomenon in a wider range of two-dimensional systems.  This first step played a major role in the quest for new topological states, as it paved the way for the generalization of the quantum Hall effect. Indeed, this work inspired studies aiming to discover topologically-ordered band structures in other spatial dimensions, and also, in physical systems satisfying time-reversal symmetry \cite{Zhang:2001}. This effort led to the prediction \cite{Kane:2005,Bernevig:2006,Fu:2007} and to the discovery \cite{Konig:2007,Roth:2009} of $Z_2$ topological insulators, which are materials featuring large spin-orbit coupling and exhibiting the quantum spin Hall effect. In fact, the transport properties offered by these materials, which currently attract the scientific community for their potential future applications, can be described by simple lattice models that are direct generalizations of the initial Haldane model \cite{Kane:2005,Qi:2008}. Besides, it was recently shown that the Haldane model could be tailored in order to create band structures presenting low-energy flat bands with non-zero Chern numbers \cite{Neupert:2011}. In such configurations, in analogy with fractional quantum Hall states occurring in strong magnetic fields, interactions can profoundly change the many-body ground state. This offers a novel route towards strongly-correlated topological liquids, the so-called  {\it fractional Chern insulators} \cite{Neupert:2011,Regnault:2011,Wu:2012,Yao:2012}. Although it played a fundamental role in the search for topological materials, the Haldane model was never reproduced in solid-state systems. Recently, it was suggested that the \emph{anomalous} QH effect could be realized and observed in HgMnTe quantum wells \cite{Liu:2008}, in silicene subjected to in-plane magnetic fields \cite{Wright:2012}, in 2D organic topological insulators \cite{Wang:2012}, and in cold-atom setups \cite{Li:2009,Stanescu:2009,Stanescu:2010,Liu:2010,Alba:2011,Goldman:2012njp,Sun:2011,Dauphin:2012}. \\

\begin{figure*}[tbp]
\begin{center}
\includegraphics[width=7.in]{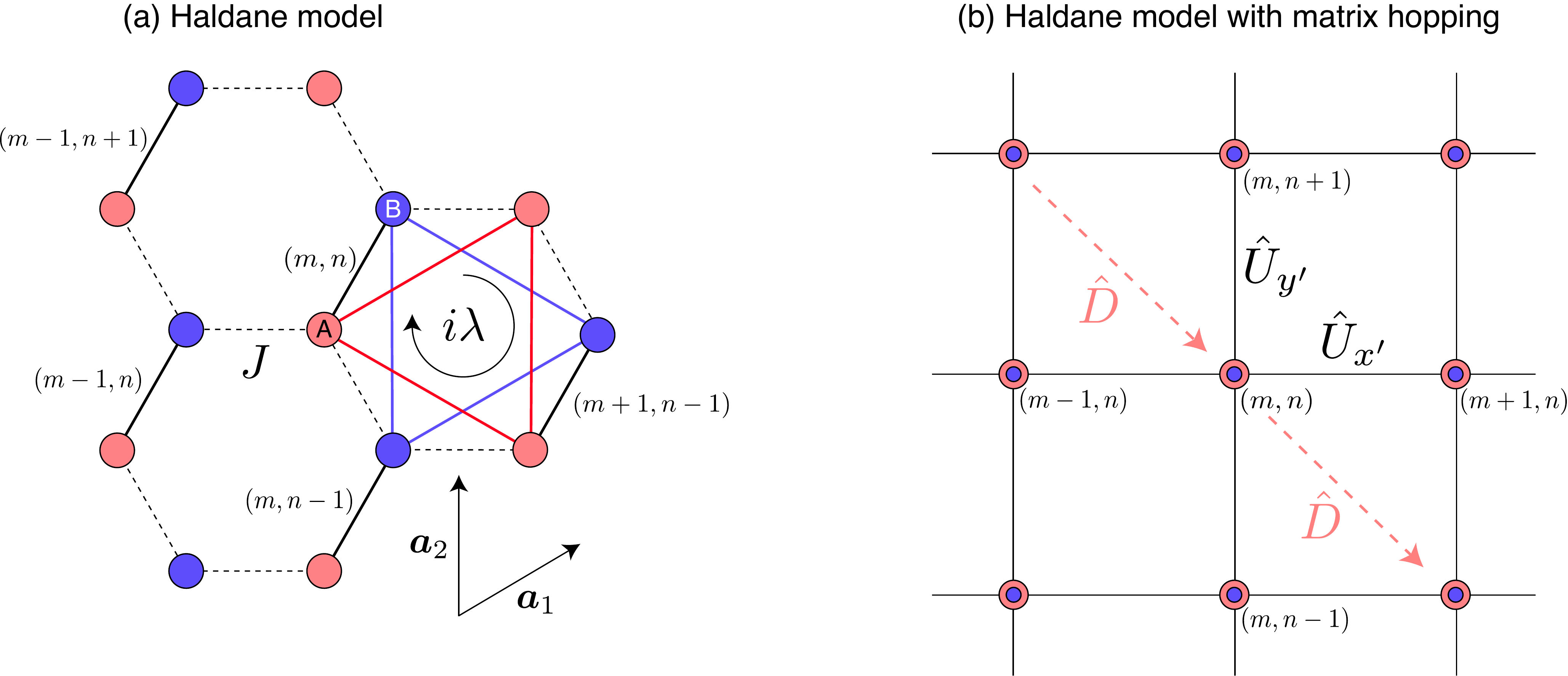}
\end{center}
\caption{(a) The Haldane model on the honeycomb lattice. The unit cells of the honeycomb are labeled by the coordinates $(m,n) \in \mathbb{Z}$. Two inequivalent sites $A,B$ belonging to the same unit cell are connected by a full black line. Standard nearest-neighbour (NN) hopping, with amplitude $J$ and denoted by dotted lines, takes place between  the  A (red) and B (blue) sites.  The complex next-nearest-neighbor (NNN) hoppings, introduced to open a quantum Hall gap, are represented by thick blue and red lines inside one honeycomb cell. The NNN tunneling factors are $+ i \lambda$ according to the orientation designated by the circular arrow, and $- i \lambda$ otherwise, i.e., they introduce a chirality in the system. (b) The same Haldane model translated into a non-Abelian square lattice, with matrix hopping operators $U_{x',y'}$. The ``undesired" diagonal hoppings $\hat{D}$ are depicted by red dotted arrows and disappear in the limit $\alpha=0$ (see  text). Note that when $\alpha=0$, this model reduces to the non-Abelian optical lattice illustrated in Fig. \ref{FIG4} (c). The modified Haldane model, corresponding to $\alpha=0$, is represented in the Appendix.}
\label{FIG1}
\end{figure*}

The original Haldane model \cite{Haldane:1988} is defined on a honeycomb lattice with real nearest-neighbor (NN) and complex next-nearest-neighbor (NNN) hoppings, see Fig. \ref{FIG1} (a). The NN hopping alone leads to a gapless energy spectrum featuring two Dirac cones familiar from graphene studies \cite{CastroNeto:2009}. The complex NNN hopping, represented by red and blue links in Fig. \ref{FIG1} (a), opens a single gap at half-filling (i.e. around $E=0$). This bulk energy gap is non-trivial \cite{Haldane:1988}: when the Fermi energy $E_{F} \approx 0$ lies in this gap, a single edge mode is populated and carries current around the system \cite{Hatsugai1993}, leading to a quantized Hall conductivity, $\sigma_H (h/e^2)=\nu =\pm 1$. From a topological point of view, the lowest energy bulk band $E (\bs k)<0$ is associated with a non-vanishing Chern number $\nu= \pm1$, which guarantees the presence of the robust and unique edge mode within the bulk gap \cite{Kohmoto:1985,Hatsugai1993,HasanKane2010}. The Chern number $\nu$ can be directly evaluated from the effective Dirac equations associated with the two Dirac cones \cite{CastroNeto:2009,Haldane:1988},
\begin{equation}
\label{eq:Chern}
\nu=\frac{1}{2}\bigl ( \txt{sign} (M_{{\bf K}^+})-\txt{sign} (M_{{\bf K}^-}) \bigr),
\end{equation}
where $M_{{\bf K}^\pm}$ are the effective masses related to the two independent Dirac points ${\bf K}^\pm$. \\

Realizing the Haldane model with cold atoms trapped in an optical lattice is attractive, as it would provide a simple playground to investigate the physics of topological (Chern) insulators. Such a proposal was described in Refs. \cite{Alba:2011,Goldman:2012njp}, where laser-induced tunneling \cite{jaksch2003a} was considered to couple two state-dependent triangular optical lattices. The simplicity of the Haldane model relies on the fact that it is a two-band model, featuring a \emph{unique} energy bulk gap and hosting a \emph{single} topological edge mode ($\nu=\pm 1$). This minimal topologically-ordered system is to be compared, for instance, to the emblematic Hofstadter model \cite{Hofstadter1976}, {\it i.e.} a tight-binding model for an electron in a uniform magnetic field  moving on a square lattice, which is characterized by a multi-band energy spectrum and hosts (infinitely many) quantum Hall phases with arbitrary $\nu \in \mathbb{Z}$ \cite{Osadchy:2001}.  In this sense, the two-band Haldane model is mathematically easier to handle than the Hofstadter model \cite{Goldman:2012njp}, and it thus  constitutes a good basis for studying the effects of interactions, in particular, in view of realizing fractional Chern insulators \cite{Neupert:2011,Regnault:2011,Wu:2012,Yao:2012,Cooper:2012} and topological Mott insulators \cite{Dauphin:2012} with cold atoms. Finally, the Haldane model constitutes the building blocks for the Kane-Mele model \cite{Kane:2005}, and therefore, its optical-lattice realization opens a possible route for the observation of the quantum spin Hall effect with cold atoms \cite{Goldman:2010prl,Beri:2011,Bermudez:2010,Mazza:2012}. \\

In this work, one exploits the fact that the honeycomb Haldane model can be mapped into a two-component square lattice with NN and NNN matrix hoppings. We demonstrate that the NNN hoppings, which are inconvenient for an optical lattice implementation, can be simply omitted: they do not contribute to the appearance of a non-trivial topological phase. This simple observation makes the realization of the Haldane model feasible with laser-assisted tunneling on a square lattice, at the price of an increased complexity: the resulting model involves non-Abelian gauge potentials \cite{osterloh2005a,goldman2009a,Goldman:2009prl}. The models stemming from this non-Abelian framework are versatile, and thus, they can be generalized to simulate an assortment of topological and Dirac-like systems \cite{Lan:2011,Barnett:2012,Mazza:2012}. 

In Section \ref{nonab}, we first  describe an extension of the scheme proposed in \cite{osterloh2005a} allowing one to realize such non-Abelian gauge potentials. In Section \ref{Haldane}, we show how these synthetic gauge potentials can be tailored in order to reproduce the original Haldane model on a square lattice. We also briefly comment on the detection of relevant signatures, based on available experimental probes. In Section \ref{piflux}, we show that similar considerations can be applied to the so-called $\pi-$flux model \cite{Hatsugai:2006,Goldman:2009prl,Neupert:2011}, which leads to identical physics with a simpler setup. 

\section{Non-Abelian gauge potentials for ultracold atoms in optical lattices}
\label{nonab}

\begin{figure*}[tbp]
\begin{center}
\includegraphics[width=7.in]{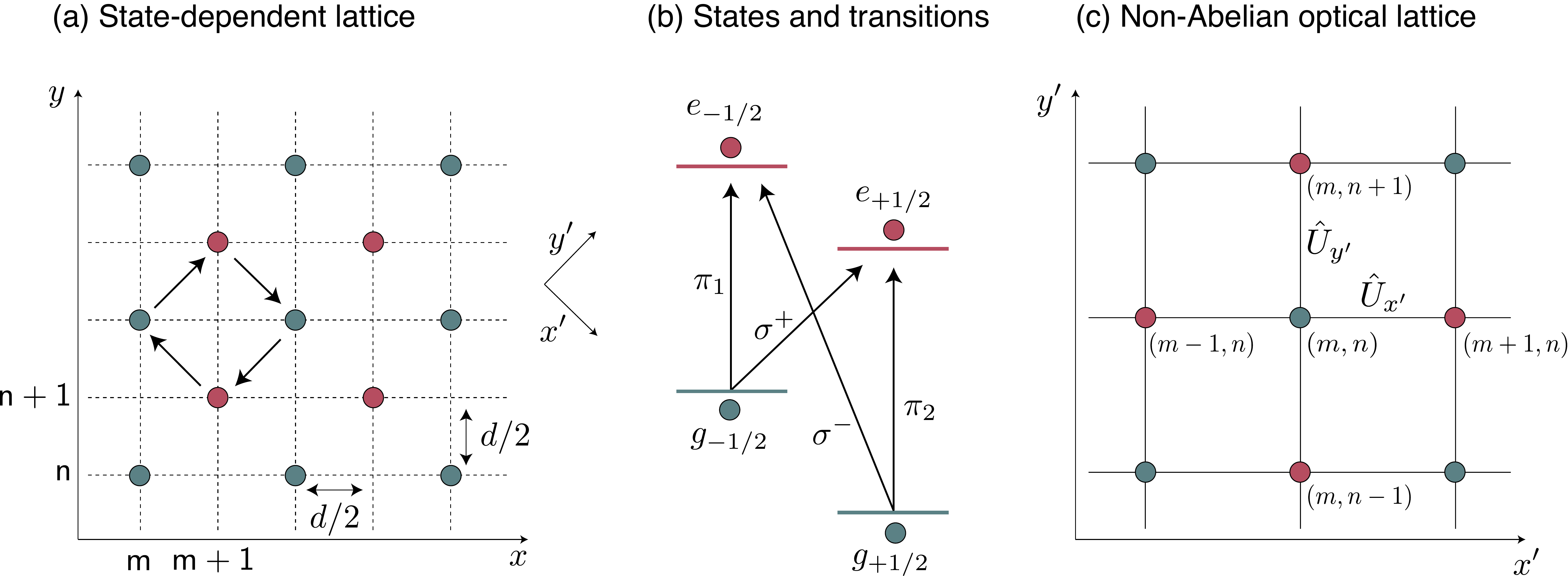}
\end{center}
\caption{(a) Sketch of the spin-dependent trapping potential for atoms with nuclear spin $1/2$. Atom in electronic state $g$, shown as green circles, (resp. $e$, shown as dark red circles) are trapped at the potential minima (resp. maxima) independently of their nuclear spin. This results in a spatial arrangement depicting a checkerboard pattern. A laser resonant on the $g-e$ transition can induce tunneling along the diagonals ${\bf e}_x'$, ${\bf e}_y'$. We show for illustration a closed trajectory around a unit cell of the checkerboard lattice. (b) Possible realization using $^{171}$Yb atoms with nuclear spin 1/2 in both $g$ and $e$ manifolds. An applied magnetic field shifts the various transitions between internal states depending on the value of the nuclear spin, allowing independent addressing of each of them (for instance, addressing the $\pi_{1,2}$ transitions independently from the $\sigma^{\pm}$ transitions). (c) The resulting ``non-Abelian"  optical lattice, with U(2) hopping operators $\hat U_{x',y'} \not\propto \hat 1_{2 \times 2}$ acting along nearest-neighboring sites. Note that we use the notation $(m,n)$ [resp. $(\mathsf{m}, \mathsf{n})$] to designate the lattice sites in the $x'-y'$ [resp. $x-y$] axis system.}
\label{FIG4}
\end{figure*}

We start by outlining an experimental scheme generating effective non-Abelian gauge fields for ultracold fermionic atoms on a lattice. The scheme is a generalization of the ones proposed by Osterloh et al. \cite{osterloh2005a}, building on the earlier proposal by Jaksch and Zoller to realize effective magnetic fields \cite{jaksch2003a}. Both proposals \cite{jaksch2003a,osterloh2005a} are designed for alkali atoms, and rely on using multilevel atoms in a spin-dependent optical lattice, {\it i.e.} a periodic trapping potential made from several different sublattices trapping different internal states. A set of laser beams drive Raman transitions that change the internal state, thereby inducing tunneling from a given sublattice to another. Crucially, the laser-assisted tunneling process is characterized by a complex matrix element, with an argument determined by the laser phase. A suitable configuration of laser beams then leads to a non-vanishing geometrical phase for an atom tunneling around a closed trajectory, which realizes an effective magnetic field. 

Implementing this scheme experimentally with fermionic alkali atoms is unfortunately plagued by spontaneous emission: the spin-dependent lattice and Raman lasers frequencies need to be set relatively close to resonance, leading to heating and losses hardly compatible with quantum gases experiments \cite{gerbier2010a}. Moreover, such spin-dependent lattices are sensitive to the magnetic moment of the atoms, leaving the protocol vulnerable to stray magnetic fields. These issues can be avoided by using alkaline earth atoms (or atoms with a similar level structure, such as Ytterbium). We outline below how the proposals must be modified to use these atoms. We first start by recalling the scheme proposed in Ref. \cite{gerbier2010a} to realize an effective magnetic field -- or equivalently an Abelian gauge structure with U(1) symmetry -- and we explicitly apply it to a checkerboard geometry, see Fig.~\ref{FIG4} (a). We then explain in a second step how this can be extended to non-Abelian configurations in the spirit of Osterloh et al. \cite{osterloh2005a}. For simplicity we consider an atom with spin $1/2$ and generations of SU(2) gauge  potentials. In principle, the method can be extended to other gauge groups but this typically requires many more lasers than the simpler U(1) or SU(2) schemes we outline here. As a result, the experimental implementation becomes significantly more involved. 

\subsection{Spin-dependent optical lattices, laser-induced tunneling and the Peierls phase}

Two-electron atoms (such as alkaline earth atoms or Ytterbium) feature two peripheral electrons that form a singlet $^1$$S_0$ ground state and several low-lying triplet states, including a long-lived metastable $^3$$P_0$ state (with a typical lifetime of tens of seconds). From now on, the notation $g= ^1$$S_0$ and  $e= ^3$$P_0$ will be used for simplicity. The ultra-narrow $g-e$ transition enables a direct coherent manipulation of the internal states without spontaneous emission. 

Consider a two-dimensional square lattice populated with atoms in the two internal states $g$  and $e$. Atoms in both states are confined to the $x-y$ plane, independently of the internal state, by a laser set near a ``magic'' wavelength where both states have equal polarizabilities \cite{gerbier2010a}. Along the $x$ and $y$ directions, two independent standing waves are applied to form a  two-dimensional square optical lattice. Here, we consider a state-\emph{dependent} optical lattice that is realized when the trapping laser is set at an ``anti-magic'' wavelength, such that the polarizability for states $g$ and $e$ are opposite: negative for $g$ atoms, which are trapped near the intensity maxima, and positive for $e$ atoms, which are trapped near the minima \cite{gerbier2010a}. We write the corresponding potentials as
\begin{align}
V_g=-V_0 \sum_{\mu=x,y} \cos ^2 (\pi \mu/d), \, V_e=+V_0  \sum_{\mu=x,y} \cos ^2 (\pi \mu/d). 
\end{align}
This results in a lattice potential with checkerboard geometry, where sites corresponding to the $e$ sublattice are centered on the cells of the $g$ sublattice (see Fig.\ref{FIG4} (a)). In the following, it will be convenient to consider a rotated frame, with axes ${\bf e}_y'=\left( {\bf e}_x+{\bf e}_y \right)/\sqrt{2}$ and ${\bf e}_x'=\left( {\bf e}_x-{\bf e}_y \right)/\sqrt{2}$  corresponding to the diagonals of the original square sublattices.

Tunneling between $g/e$ sublattices, with amplitude $J$, is  enforced by a near-resonant laser beam. As shown below, the coupling matrix element realizes the so-called ``Peierls-Luttinger substitution'' $J \rightarrow J e^{i \mathcal{A}}$, describing how a gauge potential $\mathcal{A}$ modifies the particle tunneling  $J$ (and hence, the band structure) in the tight-binding limit \cite{Hofstadter1976,Luttinger:1951,Jimenez:2012,Struck:2012}. Following Jaksch and Zoller \cite{jaksch2003a}, we assume the atoms only occupy the fundamental Bloch band and we make the tight-binding approximation.  We then obtain the effective $g-e$ hopping matrix element between two neighboring sites, located at ${\bf r}_{g}={\bf r}_{j}$ and ${\bf r}_{e}={\bf r}_{k}={\bf r}_{j}+\bs{\delta}_{jk}$:
\begin{equation}
J_{jk}=\frac{\hbar \Omega}{2} \int w_{e}^\ast({\bf r} -{\bf r}_{k})  w_{g}({\bf r} -{\bf r}_{j}) e^{i {\bf q}\cdot {\bf r}}\;{\rm d}^{2}{\bf r},  \label{JZeq}
\end{equation}
where $\Omega$ and ${\bf q}$ are the laser Rabi frequency and wavevector, respectively, and where $w_{g/e}$ denote the Wannier functions associated with each sublattice. For the checkerboard geometry, the link vectors are given by $\bs{\delta}_{jk}={\bf r}_{k}-{\bf r}_{j}={\bf e}'_{x,y}$. It is convenient to rewrite the tunneling elements as
\begin{align}
&J_{jk}=J_{\text{eff}} \, e^{i {\bf q} \cdot ({\bf r}_{j} + {\bf r}_{k})/2}=J_{\text{eff}} \, e^{i \phi_{jk}} ,  \label{JZeq2}\\
&J_{\text{eff}}=\frac{\hbar \Omega}{2} \biggl ( \int w_{k}^\ast({\bf r} -\bs{\delta}_{jk}/2)  w_{j}({\bf r} + \bs{\delta}_{jk}/2) e^{i {\bf q}\cdot {\bf r}}\;{\rm d}^{2}{\bf r} \biggr ), \nonumber
\end{align}
in which case the magnitude of the tunneling $J_{\text{eff}}=(J_{\text{eff}})^*$ is uniform over the whole lattice. We note that other choices are acceptable for the phases $\phi_{jk}$, as long as the new phases $\tilde \phi_{jk}$ satisfy the gauge-transformation relation $\tilde \phi_{jk}=\phi_{jk} + \chi ({\bf r}_{k}) -  \chi ({\bf r}_{j})$, where the function $\chi (\bs r)$ is defined uniformly over the whole lattice. The magnitude of the tunneling matrix element $J_{\text{eff}} $ is controlled by the Rabi frequency and the overlap integral between the Wannier functions $w_{g/e}$ associated with each sublattice, which have to be calculated numerically \cite{jaksch2003a}. In principle, the laser is also able to induce tunneling to next-nearest-neighbors of the composite lattice, and beyond. In practice, the matrix elements are exponentially suppressed compared to the ones describing nearest-neighbors tunneling, and negligible for realistic experimental configurations. This sets a constraint on the class of models that can be realized in this way, as we will discuss later in the paper.\\

From now on, we use a shorthand notation $(\sf{m},\sf{n})$ for the site located at $x=\sf{m} d/2, y=\sf{n} d/2$, where $\sf{m},\sf{n}$ are even for state $g$ and odd for state $e$, and where $d$ is the lattice spacing for each sublattice, see Fig. \ref{FIG4}(a). Also, we choose ${\bf q}=q \, \bs{e}_y$ for definiteness. The effective $g-e$ hopping matrix elements between a site located at ${\bf r}_{g}=(\sf{m},\sf{n})$ and its four nearest neighbors are then given by
\begin{align}
&{\bf r}_{g} \rightarrow {\bf r}_{e}= {\bf r}_{g} + {\bf e}'_{x}: \quad J_{ge}=J_{\text{eff}} \, e^{i \pi \alpha (\sf{n} - 1/2)},  \nonumber \\
&{\bf r}_{g} \rightarrow {\bf r}_{e}= {\bf r}_{g} + {\bf e}'_{y}: \quad J_{ge}=J_{\text{eff}} \, e^{i \pi \alpha (\sf{n} + 1/2)}, \nonumber \\
&{\bf r}_{g} \rightarrow {\bf r}_{e}= {\bf r}_{g} - {\bf e}'_{x}: \quad J_{ge}=J_{\text{eff}} \, e^{i \pi \alpha (\sf{n} + 1/2)},  \nonumber \\
&{\bf r}_{g} \rightarrow {\bf r}_{e}= {\bf r}_{g} - {\bf e}'_{y}: \quad J_{ge}=J_{\text{eff}} \, e^{i \pi \alpha (\sf{n} - 1/2)}, \label{JZeq3}
\end{align}
and the reversed paths are given by $J_{eg}=\bigl (J_{ge}\bigr )^*$. Here, the phase factor has been written as $ 2\pi \alpha=q d$, where $\alpha$ can be controlled from zero to a value larger than one by modifying the orientation of the coupling laser. \\

Let us note that the checkerboard lattice considered here potentially allows to induce the tunneling and the Peierls phases along \emph{both} spatial directions (in contrast with the ``column" geometry previously considered in Refs. \cite{jaksch2003a,gerbier2010a,aidelsburger2011a}). This central ingredient will be largely exploited in the following of the present proposal, see Section \ref{sectionnapotential}.

\subsection{Realizing Abelian gauge potentials: non-zero flux on the checkerboard lattice}
 First of all, we note that a particle making a loop around a unit cell of the checkerboard lattice with tunneling elements \eqref{JZeq2} does not acquire any phase factor. This is a consequence of unitarity, which imposes that the laser coupling matrix elements for  $g \rightarrow e$ transitions [see Eq. \eqref{JZeq2}] and its reversed $e \rightarrow g$ counterpart, are complex conjugates. Because of the alternating pattern of $g \rightarrow e$ and $e \rightarrow g$ links [see Fig. \ref{FIG4} (a)], the sign of the tunneling phases alternates when moving in a given direction, which indeed leads to the cancellation of the overall phase picked by a particle hopping around a unit cell of the checkerboard. Starting from the site $(\sf{m},\sf{n})$, and using Eqs. \eqref{JZeq3}, one obtains that the phase acquired along the closed path $(\sf{m},\sf{n})\rightarrow(\sf{m}+1,\sf{n}+1)\rightarrow (\sf{m}+2,\sf{n}) \rightarrow(\sf{m}+1,\sf{n}-1)\rightarrow (\sf{m},\sf{n})$ is
\begin{align}
2 \pi \Phi_{\square}&=\pi \alpha (\sf{n}+1/2) - \pi \alpha (\sf{n}-1/2) \nonumber \\
&+ \pi \alpha (\sf{n}-1/2) - \pi \alpha (\sf{n}+1/2)=0, 
\end{align}
which indicates that the effective ``magnetic" flux $\Phi_{\square}$ penetrating each plaquette is zero. This trivial flux configuration is in contrast with the ``column" geometry discussed in Refs. \cite{jaksch2003a,gerbier2010a,aidelsburger2011a}, where a staggered effective magnetic flux is generated, and it results from the higher symmetry of the checkerboard lattice. This can be cured by an additional superlattice with double period $2d$, generating a potential of the form 
\be
V^{\text{SL}}_{e,g}(x) = W_{e,g} \cos^2(\pi x/2d+ \varphi), 
\ee
which acts, a priori, differently on the $g$ and $e$ atoms. A suitable choice of the relative phase $\varphi$, which we take to satisfy the relation $\tan \varphi=-W_g/W_e$, leads to on-site energies
\begin{align}
&E_{g}-V_0 + W_g/2 +\Delta V, \quad ~\mbox{for} ~x/d= 4\sf{m},\nonumber\\
&E_{e} + W_e/2 +\Delta V, \qquad \quad \, \, ~\mbox{for} ~x/d= 4\sf{m}+1,\nonumber\\
&E_{g}-V_0 + W_g/2 -\Delta V, \quad ~\mbox{for}~ x/d= 4\sf{m}+2,\nonumber\\
&E_{e} + W_e/2 -\Delta V, \qquad \quad \, \, ~\mbox{for} ~x/d= 4\sf{m}+3.\nonumber
\end{align} 
Here $E_{e,g}$ denote the internal energies in free space, and $\Delta V =\frac{W_g W_e}{2 \sqrt{W_e^2 + W_g ^2}}$. In this potential landscape, the resonance frequencies for transitions linking neighboring sites become non-degenerate,
\begin{eqnarray}
\omega_{1}&=&\omega_{0}+ \delta V / \hbar \quad \, \, \, \, \, \mbox{for}~(4\sf{m},4\sf{n}) \rightarrow(4\sf{m}+1,4\sf{n}+1),\nonumber\\
\omega_{2}&=& \omega_{1} + 2 \Delta V/\hbar  \quad \mbox{for}~(4\sf{m}+1,4\sf{n}+1)\rightarrow(4\sf{m}+2,4\sf{n}),\nonumber\\
\omega_{3}&=&\omega_{1}\qquad \qquad \, \, \, \, \, \, \, \, \mbox{for}~ (4\sf{m}+2,4\sf{n}) \rightarrow(4\sf{m}+3,4\sf{n}+1),\nonumber\\
\omega_{4}&=&\omega_{1} - 2 \Delta V/\hbar \quad \mbox{for} ~(4\sf{m}+3,4\sf{n}+1) \rightarrow(4\sf{m}+4,4\sf{n}),\nonumber
\end{eqnarray} 
with $\hbar\omega_{0}=E_{e}-E_{g}$ the bare transition frequency and $\delta V = V_0 + (W_e - W_g)/2$.

To the state-dependent lattice, one thus applies three coupling lasers propagating along $y$. The laser at frequency  $\omega_{1}$ is chosen with a wavevector ${\bf q}=q {\bf e}_y$, and the lasers at frequencies $\omega_{1} \pm 2 \Delta V$ with the opposite wavevector ${\bf q}'=-q {\bf e}_y$. If we neglect off-resonant transitions, the alternation of the wavevectors compensates the alternation of the sign of the tunneling phases,  thereby leading to a non-zero ``magnetic" flux per plaquette that is uniform across the lattice. Starting from the site $(\sf{m},\sf{n})$, one obtains that the phase acquired along the closed path $(\sf{m},\sf{n})\rightarrow(\sf{m}+1,\sf{n}+1)\rightarrow (\sf{m}+2,\sf{n}) \rightarrow(\sf{m}+1,\sf{n}-1)\rightarrow (\sf{m},\sf{n})$ is
\begin{align}
2 \pi \Phi_{\square}&=\pi \alpha (\sf{n}+1/2) + \pi \alpha (\sf{n}+1/2) \nonumber \\
&- \pi \alpha (\sf{n}-1/2) - \pi \alpha (\sf{n}-1/2)= 2 \pi \alpha, 
\end{align}
where the parameter $\alpha=qd/2 \pi$ is interpreted as the uniform synthetic magnetic flux  (in units of the flux quantum) penetrating each plaquette. Thus, this scheme realizes the Hofstadter model \cite{Hofstadter1976},  generalizing the proposals \cite{jaksch2003a,gerbier2010a} to the case of the checkerboard geometry.

\subsection{Realizing non-Abelian gauge potentials}
\label{sectionnapotential}
This proposal can be generalized to non-Abelian gauge potentials when $g$ and $e$ now represent manifolds of degenerate states (see Fig.~\ref{FIG4} (a)-(b)). Let us consider a $2\times2$ system with states $\{ g_{-1/2},g_{+1/2}\}$ and $\{e_{-1/2},e_{+1/2}\}$, as realized in the fermionic $^{171}$Yb with nuclear spin $1/2$. In a moderate magnetic field (a few tens of Gauss), the Zeeman effect allows to distinguish between $\sigma^{\mp}$ transitions $g_{\pm 1/2}\rightarrow e_{\mp1/2}$, that change the spin projection, from $\pi_{1,2}$ transitions $g_{\pm 1/2}\rightarrow e_{\pm1/2}$ that conserve it (see Fig.~\ref{FIG4} (b)). This allows to correlate tunneling in a spatial direction with rotations in internal states and state-dependent tunneling phases. Laser-assisted tunneling along $x'$ or $y'$ is now described by $2\times2$ matrices $\hat{U}_{x'}$ and $\hat{U}_{y'}$ acting on a two-component spinor. If it is possible to arrange such that $[\hat{U}_{x'},\hat{U}_{y'}]\neq0$, this mimics a non-Abelian gauge potential ${\bf \hat{A}}$ through ${\bf \hat{U}}=e^{i \int  {\bf \hat{A}}\cdot d{\bf l}}$ \footnote{Technically, one should also verify that the tunneling matrices $\hat{U}_{x',y'}$ correspond to non trivial Wilson loops, see \cite{goldman2009a}.}. 

In order to realize $\hat{U}_{x'}\neq \hat{U}_{y'}$, one must distinguish between the tunnelings directed along  ${\bf e}_x'$ and ${\bf e}_y'$. Therefore, this requires two superlattice potentials with period $2d$ along the $x$ and $y$ axis, with different heights $\Delta V_x$ and $\Delta V_y$. This generates nine different transition frequencies differing by $0, \Delta V_x, \Delta V_y, \pm \Delta V_x \pm \Delta V_x,2\Delta V_x,2\Delta V_y$, that can be grouped into two families. One family corresponds to transitions along ${\bf e}_x'$ and the other one corresponds to transitions along ${\bf e}_y'$. 

We first consider the transitions along ${\bf e}_y'$ for illustration. Let us choose the coupling lasers propagating along $z$ (thus inducing tunneling with no relative phase for motion in the $x-y$ plane), with a linear polarization $\sigma^x$, and tuned on resonance in the absence of magnetic field. Such lasers will induce a tunneling matrix $\propto \hat{\sigma}_x$. A second set of lasers is added, with polarization $\sigma^y$, on resonance with transitions along ${\bf e}_x'$. This configuration generates a spin-orbit coupling of the form
\begin{eqnarray}
\hat{U}_{x'} \propto \hat{\sigma}_y, ~\hat{U}_{y'} \propto e^{i\phi_1} \hat{\sigma}_x,
\end{eqnarray} 
where $\phi_1$ is the relative phase of the $\sigma^y$-polarized laser with respect to the $\sigma^x$-polarized one.

More complicated tunneling matrices with complex position-dependent elements can be generated using a different laser configuration. Two specific configurations, leading to the analog of the Haldane model \cite{Haldane:1988}, will be discussed below. In principle, arbitrary tunneling matrices can be generated in this way. However, the increase in complexity makes it experimentally very challenging. The configuration described above, for instance, generates no less than $9$ different frequencies. Those frequencies are typically in the $10~$kHz range, much less than the absolute optical frequency $\omega_0$ corresponding to the bare atomic transition. As a result, they can be generated from a single laser using acoustic-optical or electro-optical modulators. This represents nevertheless a significant technological challenge for designing arbitrary tunneling matrices.

\section{The Haldane model on the square lattice}
\label{Haldane}

\subsection{Mapping the Haldane model to a non-Abelian model on a square lattice}

The idea that we expose in this Section is to propose and analyze a mapping from the original Haldane model, involving complex tunneling phases on a honeycomb lattice \cite{Haldane:1988}, into a spin-$1/2$ model defined on the square lattice, see Figs. \ref{FIG1} (a)-(b). This is possible in two dimensions, since any lattice with $N$ inequivalent sites within its unit cell can be formally labelled using: (a) two spatial coordinates $(m,n)$ describing the location of the unit cell, and (b) an additional ``pseudo-spin" index $\tau=1, \dots, N$ labeling the inequivalent sites. Since the honeycomb lattice features two inequivalent sites, denoted $A$ and $B$, it can be mapped into a square lattice hosting spin-$1/2$ objects. 

Our starting point is to describe the honeycomb lattice using the notations illustrated in Fig. \ref{FIG1} (a). In these notations \cite{Hatsugai:2006}, one uses discrete coordinates $(m,n)$ to label the unit cells of the honeycomb, where $m,n$ are integers. Here each ``site" $(m,n)$ hosts  the wave-functions $\psi_A (m,n)$ and $\psi_B (m,n)$, associated with the two inequivalent sites of the honeycomb. In other words, the honeycomb lattice can be interpreted as a square lattice with a pseudo-spin $1/2$ structure. Thus, the tight-binding Hamiltonian describing NN hopping between $A$ and $B$ neighboring sites can be mapped into an equivalent operator, which involves hopping between opposite pseudo-spin components, which are located at neighboring sites of a square lattice, see Fig. \ref{FIG1} (b). This mapping smoothly deforms the original Brillouin zone and energy bands but does not affect the physical properties, such as bulk gaps and topological phases, which are discussed in the following. \\

One considers the Haldane model, which features both NN and NNN hoppings on the honeycomb lattice (Fig. \ref{FIG1} (a)). The single-particle Schr\"odinger equations, satisfied by the wave-functions $\psi_A (m,n)$ and $\psi_B (m,n)$, take a very simple form in terms of the notations introduced above, and read 
\begin{align}
&E \psi_A (m,n)= \psi_B (m,n)+  \psi_B (m-1,n) + \psi_B (m,n-1)  \notag \\
&+ i \lambda  \biggl ( \psi_A (m+1,n) - \alpha \psi_A (m+1,n-1) + \psi_A (m,n-1) \notag \\
& - \psi_A (m-1,n) + \alpha \psi_A (m-1,n+1)
- \psi_A (m,n+1) \biggr ) , \notag \\
&E \psi_B (m,n)= \psi_A (m,n)+  \psi_A (m+1,n) + \psi_A (m,n+1) \notag \\
&+ i \lambda \biggl ( \psi_B(m,n+1) -  \psi_B (m+1,n) + \alpha \psi_B (m+1,n-1) \notag \\
&- \psi_B (m,n-1) + \psi_B (m-1,n)
- \alpha \psi_B (m-1,n+1) \biggr ) ,
\label{eq:one}
\end{align}
where the tunneling amplitude $J$ to nearest-neighbors of the honeycomb lattice is set to $J=-1$, and thus defines our unit of energy. Here, $ \lambda$ is the amplitude of the NNN hoppings on the honeycomb lattice and it corresponds to the $t_2$ parameter of the original Haldane model (or equivalently to the intrinsic spin-orbit coupling strength in the spin-1/2 Kane-Mele model \cite{Kane:2005}). We remind that the complex NNN hoppings,  which are responsible for the opening of the non-trivial topological bulk gap in this model, introduce a chirality in the system, see Fig. \ref{FIG1} (a). Note that we have included an additional parameter $\alpha \in [0,1]$ -- for reasons that will appear clear after the mapping to a square lattice -- with $\alpha=1$ in the original Haldane model. \\

Setting $\Psi (m,n)= (\psi_A (m,n), \psi_B (m,n))$, one can rewrite Eq. \eqref{eq:one} as
\begin{align}
E \Psi(m,n)&= \hat{F} \Psi (m,n)+ \hat{U}_{x'} \Psi (m+1,n) + \hat{U}_{y'} \Psi(m,n+1) \notag \\
& + \hat{D} \Psi (m+1, n-1)+\hat{U}_{x'}^{\dagger} \Psi (m-1,n)  \notag \\
&+ \hat{U}_{y'}^{\dagger} \Psi(m,n-1) + \hat{D}^{\dagger} \Psi (m-1, n+1) ,
\label{eq:two}
\end{align}
where $\hat{U}_{x'}=\hat{\sigma}_-+i \lambda   \hat{\sigma}_z $, $\hat{U}_{y'}=\hat{\sigma}_- - i \lambda  \hat{\sigma}_z  $, $\hat{F}= \hat{\sigma}_x$ and $\hat{D}= -i \alpha \lambda    \hat{\sigma}_z=-i t_{\txt{diag}} \hat{\sigma}_z$. Here, we have introduced the Pauli matrices $\hat{\sigma}_{x,y,z}$ and $\hat{\sigma}_-= (\hat{\sigma}_x -i \hat{\sigma}_y)/2$.  Equation \eqref{eq:two} describes a non-interacting two-component system, evolving on a \emph{square} lattice, see Fig. \ref{FIG1} (b). The obtained model is characterized by non-Abelian hopping operators $\hat{U}_{x',y'}$ that act along the links, but also by an onsite spin-mixing term $\hat{F}$ and diagonal hoppings $\hat{D}$ with amplitude $t_{\txt{diag}}=\alpha \lambda$, see Fig. \ref{FIG1} (b). To keep with the notations used in the preceding Section, we denote by $x',y'$ the principal axes of the square lattice.\\

Realizing non-Abelian hoppings $U_{x',y'}$ between the nearest-neighbors of a square optical lattice is a possible but difficult task, as discussed above. The most important issue is the presence of additional diagonal matrix hoppings $\hat{D}$, which will be exponentially small in a cold atom realization compared to $U_{x',y'}$ (see discussion in the preceding Section). Consequently, a natural question arises: how important is this diagonal hopping $\hat{D}$ for the obtention of the topological phase? Would the non-trivial bulk gap and corresponding edge-states survive in the limit $\alpha \rightarrow 0$? We answer this question by investigating the fate of the energy spectrum and topological order (Chern number and edge-states) as the Haldane model described by Eq. \eqref{eq:two} is reduced to the simpler model with $\hat{D}=0$. \\

In momentum space, the Hamiltonian describing our system reads
\begin{align}
H(k_{x'},k_{y'})&= {\bf d} (\bs k) \cdot {\bf \hat{\sigma}}, 
\end{align}
where ${\bf \hat{\sigma}}$ is a vector of Pauli matrices and where the vector ${\bf d} (\bs k)$ is given by
\begin{eqnarray}
{\bf d} (\bs k) &= 
\begin{pmatrix}
\cos(k_{x'})+\cos(k_{y'})+1\\
\sin(k_{x'})+\sin(k_{y'})\\
2\lambda \left[\sin(k_{y'})-\sin(k_{x'})\right] +2\alpha \lambda \sin(k_{x'}-k_{y'})
\end{pmatrix}.\nonumber
\end{eqnarray}

\begin{figure}
\begin{center}
\includegraphics[width=3.4in]{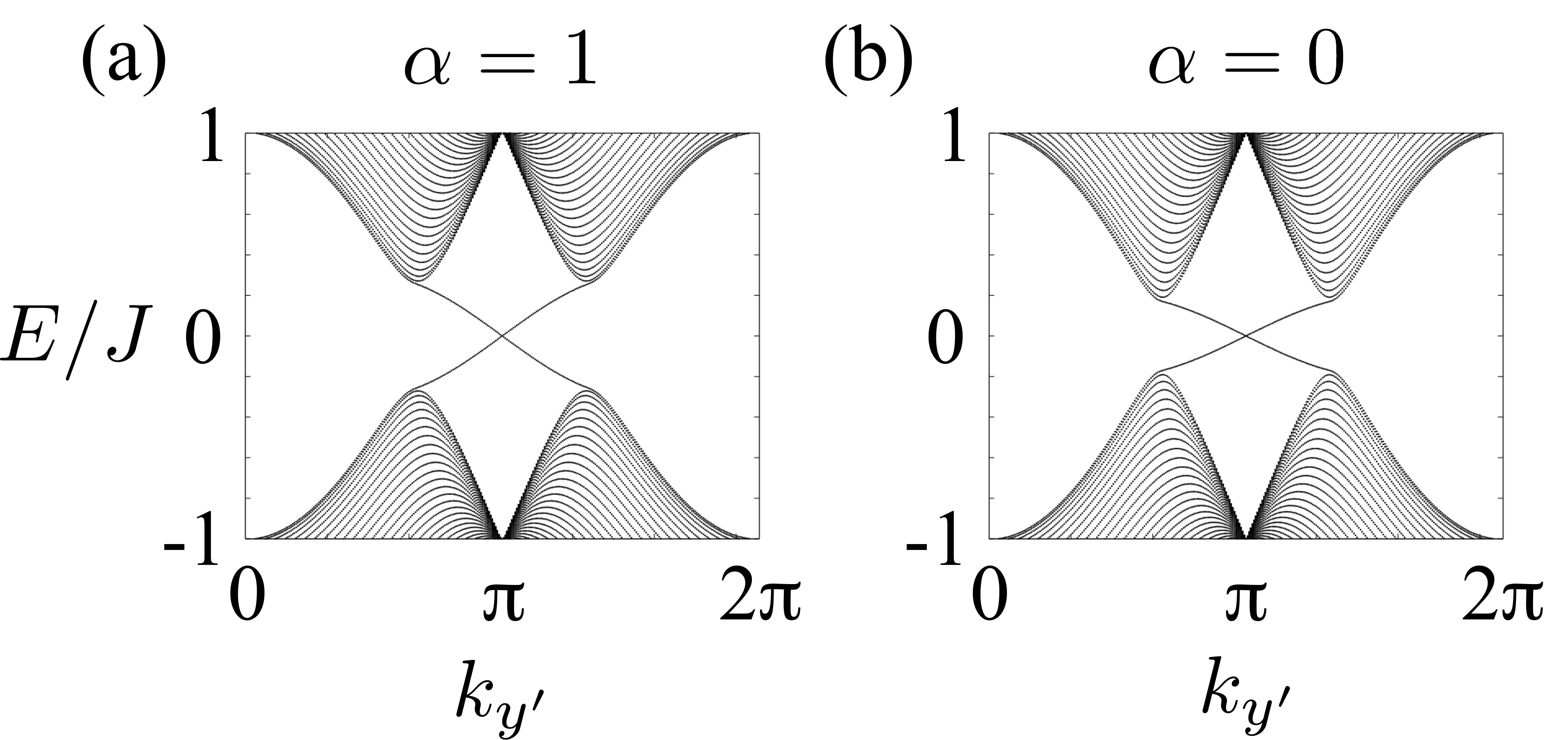}
\end{center}
\caption{Energy spectrum $E=E(k_{y'})$ for $\lambda=0.05 J$: (a) $\alpha=1$, (b) $\alpha=0$ for a cylindrical geometry aligned along $x'$. This spectrum shows the projected bulk bands $E_{\pm} (\bs{k}) \rightarrow E_{\pm} (k_{y'})$, and also reveals the presence of topological edge-states inside the bulk gap \cite{Hatsugai1993}. In the standard case $\alpha=1$, the topological bulk gap is $\Delta=6 \sqrt{3} \lambda \approx 0.5$. In the absence of the diagonal matrix hopping, i.e. $\alpha=0$, the gap is $\Delta = 4 \sqrt{3} \lambda \approx 0.34$ but survives. The energy unit is given by the tunneling amplitude $J$.}
\label{FIG2}
\end{figure}

In the limit $\lambda=0$, the energy spectrum is gapless and describes a semi-metal at half-filling. The two branches are
\begin{align}
E_{\pm}(\bs k; \lambda=0)&= \pm \mid {\bf d} (\bs k; \lambda=0)\mid,  \\
&=\pm \sqrt{3+2 \big( \cos k_{x'} + \cos k_{y'} + \cos (k_{x'}-k_{y'})\bigr )},\notag
\end{align}
and they touch at the two inequivalent Dirac points $\boldsymbol{K}^+=(2 \pi/3 , 4 \pi/3 )$,  $\boldsymbol{K}^-=(4 \pi/3 , 2 \pi/3 )$, where $E_{\pm}=0$ (or equivalently, where $d_x+id_y=0$). When $\lambda \ne 0$, a gap \begin{align}
\Delta = 2 \, \vert d_z (K^+)\vert=4 \vert \lambda \vert \sqrt{3} (1+ \frac{\alpha}{2}) 
\end{align}
opens at the Dirac points. It is well established for two-component lattice models  that the topology of the ground band is entirely characterized by the vector ${\bf d} $ parameterizing the Hamiltonian: the Chern number of the band $\nu$ is identical to the winding number of  ${\bf d}(\bs k)$, as ${\bs k}$ is varied in the first Brillouin zone \cite{Qi:2008,HasanKane2010, Alba:2011,Goldman:2012njp}. As already presented in Eq. \eqref{eq:Chern}, this winding number $\nu$ can be directly related to the effective masses associated with the two Dirac points $\bs K^{\pm}$. Here, these masses are given by
\begin{equation}
M^{\pm}=d_z( \boldsymbol{K}^\pm)=\mp 2 \lambda \sqrt{3} (1+ \frac{\alpha}{2}),
\end{equation}
as can be deduced by developing the Hamiltonian $H(k_{x'},k_{y'})$ in the vicinity of the Dirac points $\bs K^{\pm}$. The effective masses have different signs, which according to Eq. \eqref{eq:Chern}, leads to a non-zero Chern number $\nu=\pm 1$. Moreover these masses are non-vanishing in the limit $\alpha \rightarrow 0$. Therefore the bulk energy gap opened by the perturbation $\lambda$,  stemming from the complex Haldane NNN hopping, survives for any value of $\alpha$ (including the extreme value $\alpha=0$). Since the masses $M^{\pm}$ preserve their sign all along the transformation $\alpha=1 \rightarrow \alpha= 0$, the Chern number in Eq. \eqref{eq:Chern} remains constant and non-trivial for any value of $\alpha$. The latter remark can also be formulated in the following manner: since the bulk gap is preserved during this transformation, the topological Chern number characterizing the bulk bands is unaffected \cite{Kohmoto:1985}. This result is further illustrated in Fig. \ref{FIG2}, which shows the energy spectrum for the two limiting cases $\alpha=0$ and $\alpha=1$. In Figs. \ref{FIG2} (a)-(b), the spectrum was obtained using a cylindrical geometry aligned along $x'$: it shows the projected bulk bands $E_{\pm} (\bs{k}) \rightarrow E_{\pm} (k_{y'})$ and the topological edge-states inside the bulk gap \cite{Hatsugai1993}. This result demonstrates that the diagonal hopping induced by the operator $\hat{D}= -i \alpha \lambda    \hat{\sigma}_z$ in Eq. \eqref{eq:two} does not play any role in the realization of topological insulating phases and can thus be omitted for the sake of experimental feasibility. 

To summarize, the physics of the Haldane model can be explored with an optical square lattice described by the simplified tight-binding Hamiltonian 
\begin{align}
&\mathcal{H}= -J \sum_{m,n} c_{m+1,n}^{\dagger} \hat{U}_{x'} c_{m,n}+ c_{m,n+1}^{\dagger} \hat{U}_{y'} c_{m,n}\notag \\ 
& \qquad +c_{m-1,n}^{\dagger} \hat{U}_{x'}^{\dagger}  c_{m,n}  + c_{m,n-1}^{\dagger} \hat{U}_{y'}^{\dagger}  c_{m,n}+ c_{m,n}^{\dagger} \hat{F} c_{m,n},  \notag \\
&\hat{U}_{x'}=\hat{\sigma}_-+i \lambda   \hat{\sigma}_z , \quad \hat{U}_{y'}=\hat{\sigma}_- - i \lambda  \hat{\sigma}_z  , \quad \hat{F}= \hat{\sigma}_x,  \label{eq:H1}
\end{align}
where the two-component operator $c_{m,n}^{\dagger}$ creates a particle at site $(m,n)$ and where the  spin-1/2 structure is implicit. We note that this system is formally similar to the  HgMnTe quantum wells model proposed in \cite{Liu:2008,Qi:2008} to realize the anomalous quantum Hall effect.

\subsection{Implementation}

We now discuss how the Hamiltonian Eq.~(\ref{eq:H1}) can be implemented using the methods discussed in Section \ref{nonab}. We start by the terms $\hat{U}_{x'}=+i \lambda   \hat{\sigma}_z$ and $\hat{U}_{y'}'=-i \lambda   \hat{\sigma}_z$. First, we consider  transitions corresponding to tunneling along  the $x'$ direction. For each value of the nuclear spin ($+1/2$ or $-1/2$), one can apply laser-assisted tunneling with $\pi$ polarization. The lasers propagate along $z$ (thus inducing no tunneling phase), with a polarization along $y$ parallel to the direction of an applied magnetic field large enough to split the different transitions. Since the transition frequencies for $g_{1/2} \rightarrow e_{1/2}$ and $g_{-1/2} \rightarrow e_{-1/2}$ are different, they must be addressed separately with independent lasers, see Fig. \ref{FIG4} (b). We choose a phase of $\pm \pi/2$ for the two lasers driving the two $\pi$ transitions for nuclear spins $\pm1/2$. This gives a tunneling matrix proportional to
\begin{eqnarray}\nonumber
\hat{U}_{x'} &\propto \begin{pmatrix}
i &0\\
0&-i
\end{pmatrix} =i \hat{\sigma}_z.
\end{eqnarray} 
Another set of lasers can be used to generate tunneling operators along  ${\bf e}_y'$, also driving $\pi$ transitions for nuclear spins $\pm 1/2$ with phases $\mp \pi/2$. This generates a tunneling matrix
\begin{eqnarray}
\hat{U}_{y'} &\propto -i \hat{\sigma}_z.
\end{eqnarray}

Next, we consider additional coupling lasers propagating along $+{\bf e}_y$, with $\sigma^-$ polarization, thus generating a tunneling matrix $\propto \hat{\sigma}_-$.  For instance, the tunneling matrix element for the transition linking site $\bs{r}_g=(m,n)$ to $\bs{r}_e={\bf r}_{g}+{\bf e}_x'=(m+1,n)$  is then
\begin{eqnarray}
\langle m+1,n \mid \hat{U}_{x'} \mid m,n \rangle &\propto e^{i \phi_{-} (m,n)} \hat{\sigma}_-=e^{i \pi  \sf{n} \, \alpha} \hat{\sigma}_- , 
\end{eqnarray} 
where we used Eq. \eqref{JZeq}, and where $\sf{n}$ designates the $y$ coordinate of the site $(m,n)$, which is here labelled according to the $x'-y'$ axis system, see Fig. \ref{FIG4}. We note that each phase $\phi_{-} (m,n)$ is driven by a specific laser, corresponding to the different transition frequencies offered by the superlattice potential. By adjusting the relative phases of these lasers, one can annihilate the undesired phases $\phi_{-} (m, n)=0$ uniformly on the lattice, and thus realize constant tunneling operators $\hat{U}_{x',y'} \propto \hat{\sigma}_-$ as in Eq.~(\ref{eq:H1}).  Combining the lasers generating the $\pi$ and $\sigma^-$ transitions, one obtains the required NN tunneling terms in Eq.~(\ref{eq:H1}). Finally, the on-site term $ \hat F \propto \hat{\sigma}_x$ can be generated by an additional radio frequency field resonant  at the Larmor frequency, acting uniformly on all sites \footnote{Note that with our choice of phases, the phase of this radio frequency field plays the role of the phase reference, to which the phases of the lasers inducing tunneling process can be locked.}. \\


Although the scheme seems directly feasible, it is rather complex due to the large number of transition frequencies involved. In the following Section \ref{piflux}, we present an equivalent model that leads to a simpler implementation.

\subsection{Detection}

The Haldane-like optical lattice described above is characterized by the following physical properties: 
\begin{itemize}
\item a massless Dirac-like spectrum for $\lambda =0$, 
\item a massive Dirac-like spectrum for small $\lambda \ll J$, 
\item a largely gapped spectrum for $\lambda \sim J$, 
\item Chern insulating phases for $\lambda \ne 0$. 
\end{itemize}

The Dirac-like spectrum can be experimentally detected along the lines of the recent experiments \cite{Tarruell:2011,Uehlinger:2012} (see also Ref. \cite{Lim:2012}). In particular, the techniques developed in these experiments could identify the opening of the bulk gap in the vicinity of the Dirac points, and thus, evaluate the amplitude of the mass terms $M^{\pm} \ne 0$. We point out that the detection of a Dirac spectrum in our square optical lattice setup would already be an interesting signature of the non-trivial synthetic gauge potential \cite{Goldman:2009prl}.\\

\begin{figure}[tbp]
\begin{center}
\includegraphics[width=3.5in]{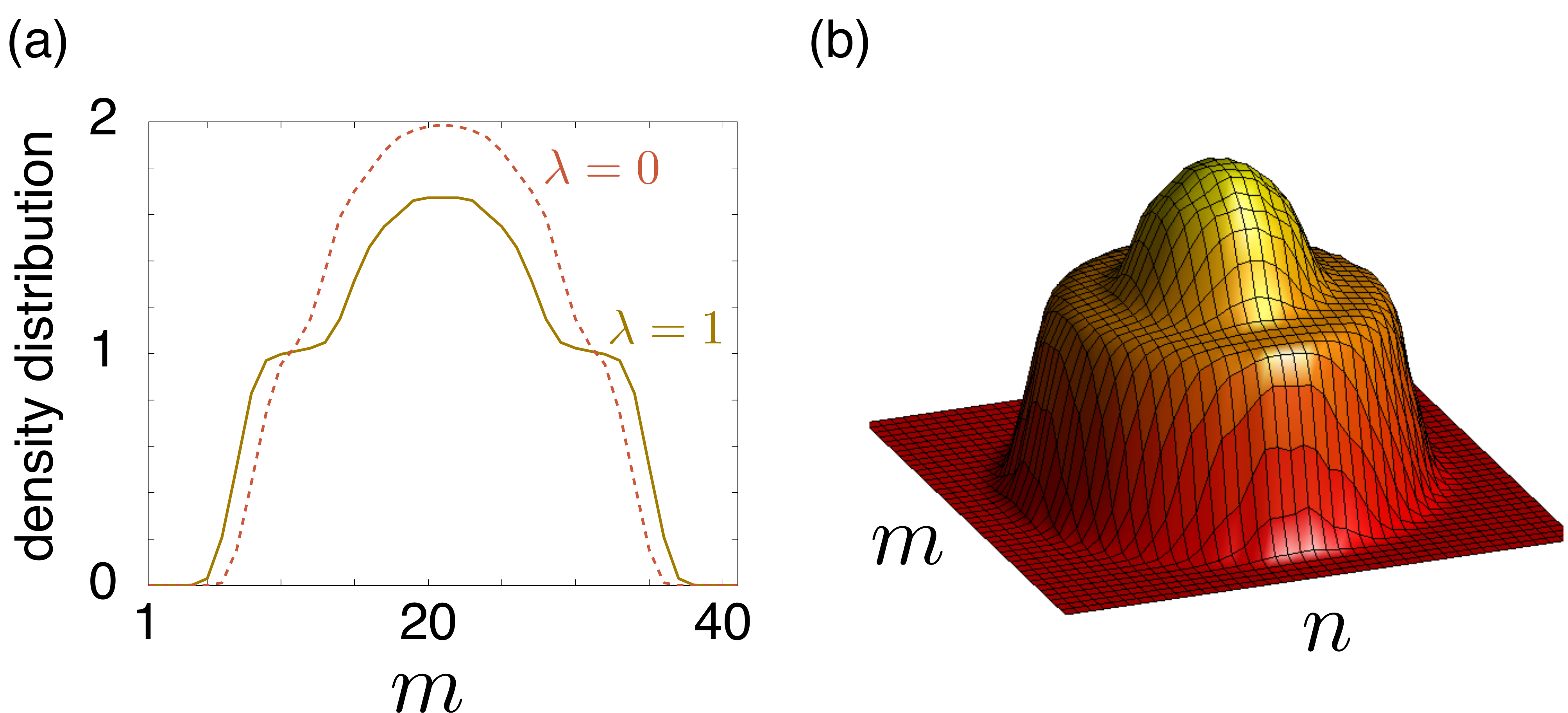}
\end{center}
\caption{(a) The spatial density for the Haldane-like optical lattice described by Eq. \eqref{eq:H1}, for $\lambda=0$ and $\lambda=1$. The atoms are trapped by a harmonic potential $V(r)= V_0 (r/r_0)^2$, with $V_0=8J$ and $r_0=17 a$, and the lattice size is $40 \times 40$. The density was obtained by setting the Fermi energy to the value $E_{\text{F}}=3J$ (corresponding to the maximum of the highest bulk band for $\lambda=0$). In this figure the $n$ coordinate is chosen to be fixed at the center of the trap. (b) Same as (a) for $\lambda=1$, but represented in the 2D plane $x'-y'$. Note the clear plateau for $\lambda=1$, which indicates the opening of a spectral gap \cite{gerbier2010a}.}
\label{FIGdensity}
\end{figure}

The large gap opened by the terms proportional to $\lambda$ could be directly detected through in-situ imaging. Indeed, this spectral gap directly manifests itself in the spatial density of the atomic cloud \cite{gerbier2010a}. We show the corresponding density for $\lambda=0$ and $\lambda=1$ in Fig. \ref{FIGdensity}, for a system trapped by an additional harmonic potential $V(r)= V_0 (r/r_0)^2$: the clear plateau depicted by the density when $\lambda=1$ is a direct signature of the gap opening. \\

The most efficient way to identify the formation of a Chern insulating phase would be to detect the presence of robust and chiral edge modes, whose energies are located within the bulk gap discussed above. This could be performed by deforming the atomic cloud and imaging its time-evolving spatial density \cite{Goldman:2012dynamics}, or using light scattering methods \cite{Goldman:2012prl,Goldman:2012special}. In principle, measuring the Chern number $\nu$ could be performed through spin-resolved momentum density measurements \cite{Alba:2011,Goldman:2012njp}, Bloch oscillations \cite{Price:2012,Abanin:2012} or hybrid time-of-flight images \cite{Wang:2013}.

\section{The {\large $\bs \pi$}-flux model}
\label{piflux}
\begin{figure*}
\begin{center}
\includegraphics[width=7.in]{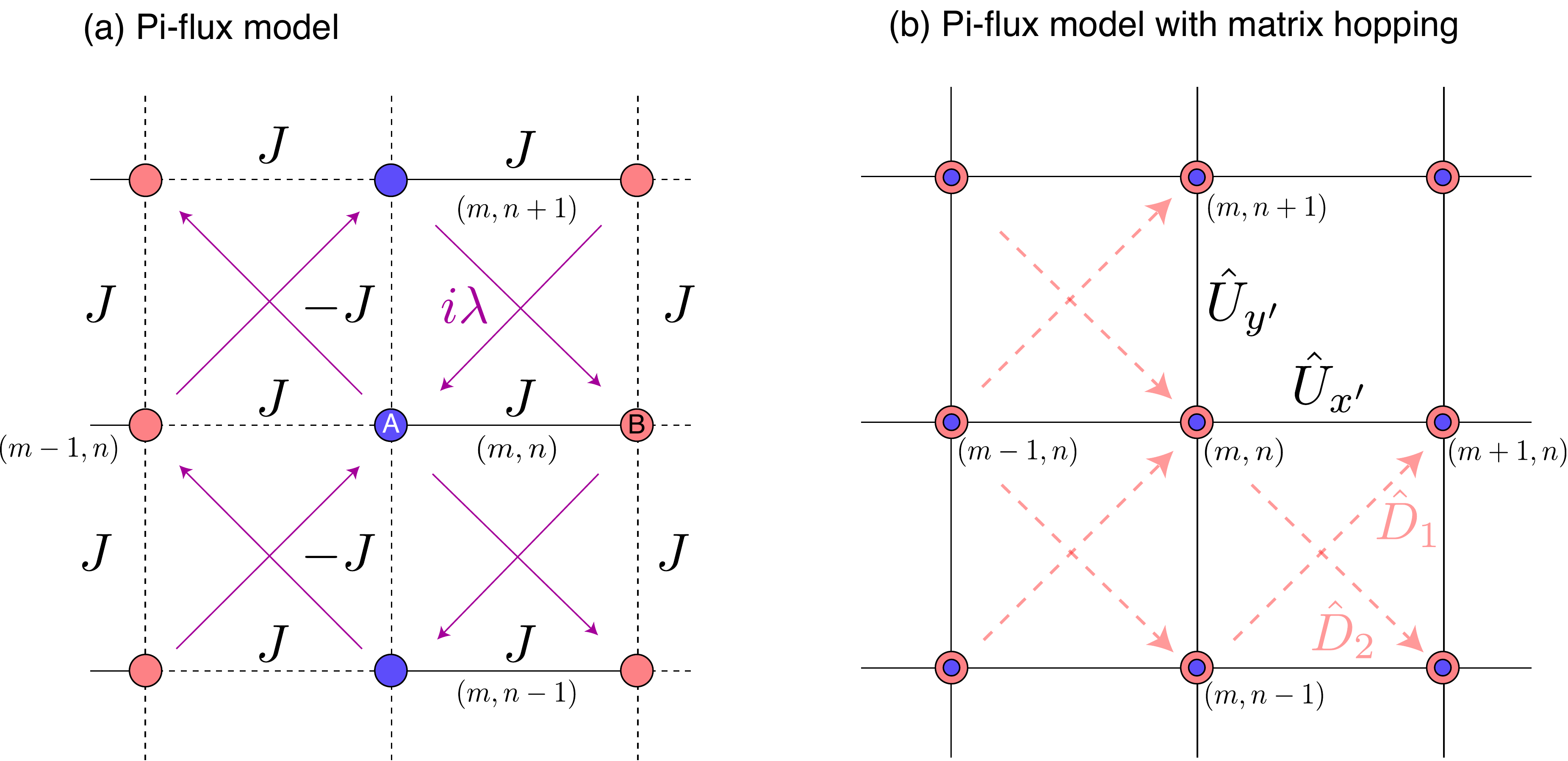}
\end{center}
\caption{(a) The $\pi-$flux model on the square lattice. The unit cells of the $\pi-$flux lattice, with inequivalent sites $A,B$, are labeled by the coordinates $(m,n) \in \mathbb{Z}$. Two inequivalent sites $A,B$ belonging to the same unit cell are connected by a full black line. Standard nearest-neighbour (NN) hoppings are characterized by the tunneling factors $\pm J$. The complex next-nearest-neighbor (NNN) hoppings, with tunneling factor $+ i \lambda$, are represented by purple dotted arrows (the hoppings with opposite factor, $- i \lambda$, correspond to the reversed paths). The chirality introduced by the NNN terms potentially results in anomalous quantum Hall phases. (b) The same $\pi-$flux model translated into a non-Abelian square lattice, with matrix hopping operators $U_{x',y'}$. The ``undesired" diagonal hoppings $\hat{D}_{1,2}$ are depicted by red dotted arrows and disappear in the limit $\alpha=0$ (cf. text). Note that when $\alpha=0$, this model reduces to the non-Abelian optical lattice illustrated in Fig. 1 (c). The modified $\pi$-flux model, corresponding to $\alpha=0$, is represented in the Appendix.}
\label{FIG3}
\end{figure*}

The physics of the honeycomb Haldane model can be alternatively studied by considering non-trivial hoppings on the $\pi$-flux model \cite{Hatsugai:2006,Goldman:2009prl,Neupert:2011}. The latter is illustrated in Fig. \ref{FIG3} (a) and is characterized by  tunneling amplitudes $\pm J$ along alternate columns. In the absence of additional hoppings, this system exhibits a gapless spectrum with two Dirac cones, and it is therefore physically equivalent to the honeycomb lattice \cite{Hatsugai:2006}. In order to open a non-trivial topological gap, with Chern number $\nu=\pm 1$, complex diagonal hoppings are required, see the purple arrows in Fig. \ref{FIG3} (a). As for the original honeycomb model discussed in the previous Section, these chiral and complex NNN hoppings highly diminish the feasibility of this Haldane-like model with cold atoms trapped in optical lattices. Note that the $\pi-$flux model without NNN hoppings is equivalent to the Hofstadter model with the magnetic flux set to $\Phi=1/2$ \cite{lim2008a,gerbier2010a}.

In this Section, we follow the same strategy as before: we consider the two-component wave-function $\Psi (m,n)=\bigl (\psi_A (m,n), \psi_B (m,n) \bigr)$, which describes particles on alternate columns, see the blue ($A$) and red ($B$) sites in Fig. \ref{FIG3} (a). In these notations, the single-particle Schr\"odinger equation satisfied by $\Psi (m,n)$ in this model then takes the form
\begin{align}
E \Psi(m,n)&= \hat{F} \Psi (m,n)+ \hat{U}_{x'} \Psi (m+1,n) + \hat{U}_{y'} \Psi(m,n+1) \notag \\
&+ \hat{D}_1 \Psi (m+1, n+1)+ \hat{D}_2 \Psi (m+1, n-1) \notag \\
&+\hat{U}_{x'}^{\dagger} \Psi (m-1,n) + \hat{U}_{y'}^{\dagger} \Psi(m,n-1) \notag \\
&+ \hat{D}^{\dagger}_1 \Psi (m-1, n-1)+ \hat{D}^{\dagger}_2 \Psi (m-1, n+1) ,
\label{eq:Hpiflux}
\end{align}
where $\hat{U}_{x'}=\hat{\sigma}_- $, $\hat{U}_{y'}=-\hat{\sigma}_z - i \lambda  \hat{\sigma}_x  $, $\hat{F}= \hat{\sigma}_x$ and $\hat{D}_{1,2}= \pm i \alpha \lambda    \hat{\sigma}_-$. Therefore, the model is translated into a non-Abelian square lattice, with direct tunneling operators $\hat U_{x',y'}$, an onsite term $\hat{F}$, and two diagonal hopping matrices  $\hat{D}_{1,2}$, see Fig. \ref{FIG3} (b). \\


The $\pi-$flux model can be analyzed in the same way as in the preceding Section, with identical conclusions. In the limit $\lambda=0$, the energy spectrum is given by
\begin{equation}
E_{\pm}(\lambda=0)=\pm \sqrt{2+2 \big( \cos k_{x'} + 2 \cos^2 k_{y'} \bigr )},
\end{equation}
with two independent Dirac points at $\boldsymbol{K}^+=(\pi , \pi/2 )$ and $\boldsymbol{K}^-=(\pi , 3 \pi/2 )$. In the vicinity of the Dirac points, the effective masses are given by
\begin{equation}
M^{\pm}=\pm 2 \lambda  (1+\alpha).
\end{equation}
As a result, the Chern number $\nu=\pm 1$ remains non-zero even without diagonal hoppings ($\alpha=0$), as found previously.\\

The $\pi-$flux quantum Hall model can thus be explored by realizing the tight-binding Hamiltonian 
\begin{align}
&\mathcal{H}= -J \sum_{m,n} c_{m+1,n}^{\dagger} \hat{U}_{x'} c_{m,n}+ c_{m,n+1}^{\dagger} \hat{U}_{y'} c_{m,n}\notag \\
&\qquad +c_{m-1,n}^{\dagger} \hat{U}_{x'}^{\dagger}  c_{m,n} + c_{m,n-1}^{\dagger} \hat{U}_{y'}^{\dagger}  c_{m,n}+ c_{m,n}^{\dagger} \hat{F} c_{m,n},\notag \\
&\hat{U}_{x'}=\hat{\sigma}_- , \qquad \hat{U}_{y'}=-\hat{\sigma}_z - i \lambda  \hat{\sigma}_x    , \qquad \hat{F}= \hat{\sigma}_x.\label{pifluxham}
\end{align}
In fact, this is the simplest scheme, exploiting non-Abelian hopping operators on a square lattice, which leads to a topological (Chern) insulating phase. Indeed, the hopping operators $\hat{U}_{x',y'}$ should necessarily contain a term proportional to a Pauli matrix, $\hat{U}_{x',y'} \propto \hat \sigma_{\mu,\nu}$, in order to induce the Dirac spectrum. Then, an additional term $\propto \lambda  \hat \sigma_{\kappa}$ should be added to at least one of these operators in order to open the topologically non-trivial gap. This minimal configuration is realized by the operators $\hat{U}_{x',y'}$ in Eq. \eqref{pifluxham}. \\

From an experimental point of view, this scheme is significantly simpler to implement than the one previously exposed in Section \ref{Haldane}, due to the fact that it features a laser-induced tunneling phase \emph{in the $y'$ direction only}, and thus, it does not require a superlattice along $x$. This demands one $\sigma^-$ coupling laser that drives tunneling along $x'$, another laser with polarization $\sigma_x$ to drive the $y'$ tunneling, and a pair of lasers driving $\pi$ transitions to generate the term $\propto \hat \sigma_z$ with the proper relative phase; each of these lasers should also have three`` sidebands" to drive all transitions introduced by the superlattice. This is not as complex as it seems at first sight, since the required sidebands can be generated by cascading modulators (one generating sidebands corresponding to different Zeeman transitions, and another one, corresponding to the different transitions introduced by the superlattice). Nevertheless, it is fair to say that this represents an important technical challenge to realize this model experimentally.

\section{Conclusion}

In conclusion, we have described an experimental scheme to generate non-Abelian gauge potentials for cold atoms in a square optical lattice. The scheme generalizes the proposals of Refs. \cite{jaksch2003a,osterloh2005a} originally formulated for alkali atoms, to two-electron atoms (alkaline earth atoms, Ytterbium, Erbium, ...), and requires additional superlattice potentials to generate finite flux per elementary cell of the lattice \cite{gerbier2010a}. We have shown that such arrangements can be used to generate topological phases with non-zero Chern number. We have detailed two examples of microscopic models exhibiting such phases, a first one based on an explicit mapping from the Haldane honeycomb lattice  to a multicomponent model defined on the square lattice with non-Abelian tunneling operators, and a second one built directly on the square lattice (the so-called ``$\pi$-flux" model). The latter model is slightly simpler, but its experimental implementation would still represent a considerable challenge due to the relatively large number of laser frequencies that need to be controlled precisely. As explained before, the different transition frequencies can be generated from a single laser using frequency modulators, a standard tool in optoelectronics. As a result, we believe that the experimental implementation is challenging but possible. Besides the models described here, generating non-Abelian gauge fields on a lattice, can simulate many different systems, including lattice-gauge-theory models \cite{osterloh2005a} and spin-orbit coupled gases \cite{Kane:2005}, with complete freedom over the choice of the microscopic Hamiltonian \cite{Lan:2011,Barnett:2012,Mazza:2012}. This is in contrast with Raman-coupled bulk systems, for example, where the form of the spin-orbit coupling is constrained \cite{Lin:2011}. We also stress that although the laser-coupled honeycomb lattice described in Refs. \cite{Alba:2011,Goldman:2012njp} should be easier to implement experimentally, it does not offer the versatility of the non-Abelian optical lattices discussed in this work. As a result, the system described in this paper has the potential to realize other kinds of topological insulating phases, such as, for instance, the $Z_2$ topological insulators characterized by a topological invariant $\nu_{Z_2}=\pm 1$ and exhibiting the quantum spin Hall effect \cite{Kane:2005,Goldman:2010prl,Beri:2011,Bermudez:2010,Mazza:2012,Mei:2012,Cocks:2012,Goldman:2012epl,Hauke:2012}.

\paragraph*{Acknowledgments} We acknowledge discussions with J. Beugnon, J. Dalibard, S. Nascimb\`ene, G. Juzeli\={u}nas, I. B. Spielman, A. Dauphin, A. Bermudez, and E. Anisimovas. This work was supported by the FRS-FNRS (Belgium), ERC AdG QUAGATUA and StG Manybo, Spanish MINCIN Grant TOQATA (FIS2008-00784 TOQATA), and the Emergences program from Ville de Paris.

\section*{Appendix: The non-Abelian optical lattice and the modified Haldane/ $\pi$-flux models}
 
\begin{figure*}
\begin{center}
\includegraphics[width=7.in]{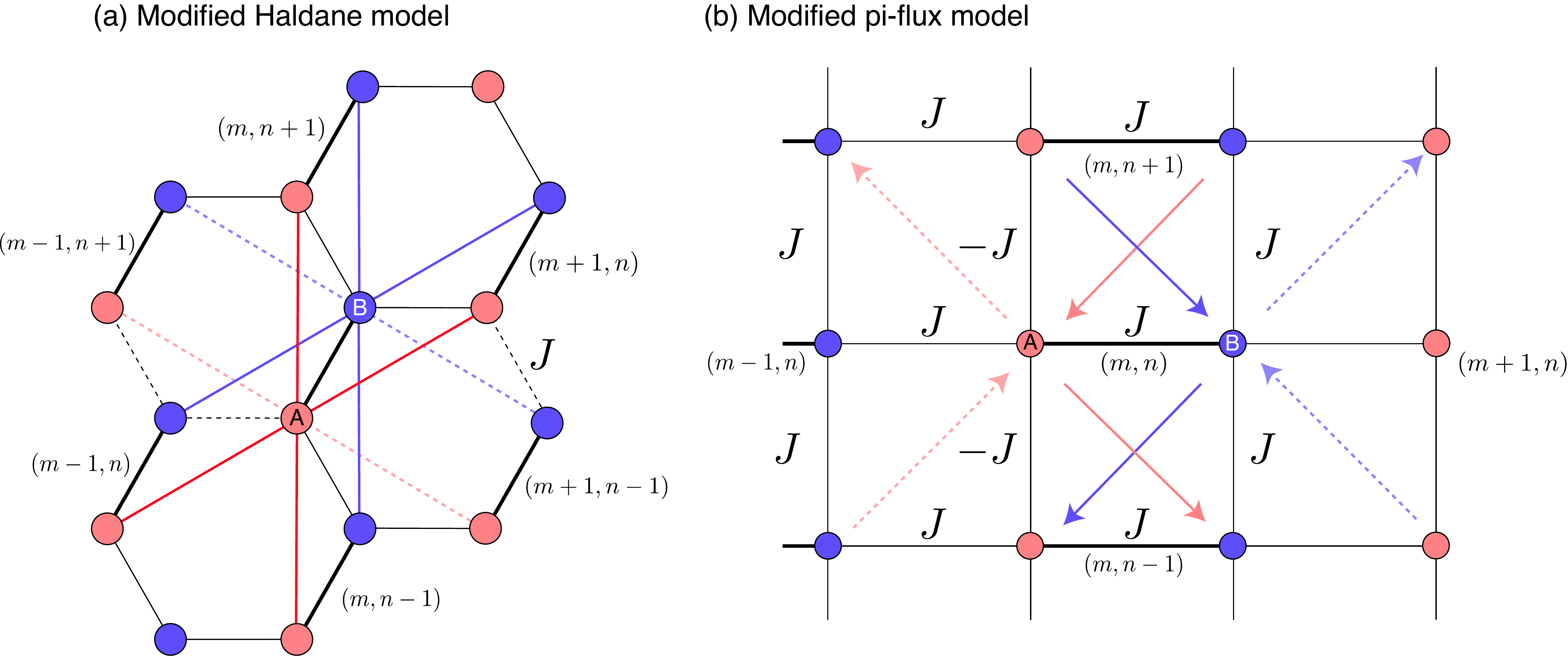}
\end{center}
\caption{(a) The modified Haldane model obtained by setting $\alpha=0$ in Eqs. \eqref{eq:one}--\eqref{eq:two}. Here we only represent the NNN hopping terms involving the two inequivalent lattice sites $A,B$ of the central unit cell $(m,n)$. The ``missing" NNN hopping terms are represented by red and blue dotted lines. (b) The modified $\pi$-flux model obtained by setting $\alpha=0$ in Eq. \eqref{eq:Hpiflux}. These pictures are to be compared with the original Haldane and $\pi$-flux models illustrated in Fig. \ref{FIG1} (a) and \ref{FIG3} (a), respectively.}
\label{FIGMissing}
\end{figure*}

 In Section \ref{Haldane}, we considered a modified version of the Haldane model that was obtained by taking the limit $\alpha \rightarrow 0$, namely, by neglecting the NNN hopping terms that led to non-zero diagonal hopping operators $\hat{D}$ in the non-Abelian model described by Eq. \eqref{eq:two}. This modification was motivated by the fact that the model with $\hat{D} \rightarrow 0$ could be realized in square optical lattices with non-Abelian hopping operators acting along NN links. To be complete, we represent in Fig.\ref{FIGMissing}(a) the NNN hopping terms of the original Haldane model that have been omitted in the transformation leading to the cold-atom model described by Hamiltonian \eqref{eq:H1}. We represent in Fig. \ref{FIGMissing} (b) the modified $\pi$-flux model, leading to the Hamiltonian \eqref{pifluxham}. 

\end{document}